\documentclass[preprintnumbers,article,amsmath,amssymb,floatfix,10pt,prd,onecolumn,
superscriptaddress,nofootinbib]{revtex4}
\usepackage[colorlinks=true, pdfstartview=FitV, linkcolor=blue, citecolor=red, urlcolor=magenta]{hyperref}
\usepackage{bbm}
\usepackage{amsfonts}
\usepackage{mathrsfs}
\usepackage{latexsym}
\usepackage{epsfig}
\usepackage{epstopdf}
\usepackage{epstopdf}
\usepackage{graphicx}
\usepackage{amssymb}
\usepackage{amsmath}
\usepackage{dcolumn}
\usepackage{bm}
\usepackage{color}
\usepackage{comment}
\usepackage{xcolor}
\begin{document}
\title{\bf Effects of Variable Equations of State on the Stability of Nonlinear Electrodynamics Thin-Shell Wormholes}
\author{Faisal Javed}
\email{faisaljaved.math@gmail.com}\affiliation{Department of
Mathematics, Shanghai University, Shanghai, 200444, Shanghai,
People's Republic of China}

\author{G. Fatima}
\email{ghulamfatima.math@gmail.com}\affiliation{Mathematics,
University of Management and Technology, Johar Town Campus, Lahore
54782, Pakistan}

\author{G. Mustafa}
\email{gmustafa3828@gmail.com}\affiliation{Department of
Mathematics, Shanghai University, Shanghai, 200444, Shanghai,
People's Republic of China}

\author{Ali {\"O}vg{\"u}n}
\email{ali.ovgun@emu.edu.tr}\homepage{https://aovgun.weebly.com}\affiliation{Physics
Department, Eastern Mediterranean University, Famagusta, North
Cyprus via Mersin 10, Turkey.}

\begin{abstract}
This paper explores the role of nonlinear electrodynamics on the
stable configuration of thin-shell wormholes formulated from two
equivalent geometries of Reissner-Nordstr\"om black hole with
nonlinear electrodynamics. For this purpose, we use cut and paste
approach to eliminate the central singularity and event horizons of
the black hole geometry. Then, we explore the stability of the
developed model by considering different types of matter
distribution located at thin-shell, i.e., barotropic model and
variable equations of state (phantomlike variable and Chaplygin
variable models). We use linearized radial perturbation to explore
the stable characteristics of thin-shell wormholes. It is
interesting to mention that Schwarzschild and Reissner-Nordstr\"om
black holes show the unstable configuration for such type of matter
distribution while Reissner-Nordstr\"om black hole with nonlinear
electrodynamics expresses stable regions. It is found that the
presence of nonlinear electrodynamics gives the possibility of a
stable structure for barotropic as well as variable models. It is
concluded that stable region increases for these models by
considering higher negative values of coupling constant $\alpha$ and
the real constant $n$.\\\\\\
\textbf{Keywords}: Thin-shell wormholes; Nonlinear electrodynamics;
Israel formalism; Stability analysis.
\end{abstract}

\maketitle

\date{\today}


\section{Introduction}

In 1934, Born and Infield \cite{1} proposed nonlinear
electrodynamics to ensure that the point-like charge self-energy is
finite. It has been found that the effective action for the open
string terminating on $D$-branes may be described perfectly in the
nonlinear form until the 1980s \cite{2}. These enormous discoveries
greatly increase the researcher's desire to explore nonlinear
electrodynamics in the context of cosmology \cite{5}. It is then
discovered that the initial Big-Bang singularity may be avoided if
the early cosmos is greatly influenced by the nonlinear
electromagnetic field. This is not the only way to produce a phase
of cosmic inflation. In addition, the nonlinear string technique to
electrodynamics has been used by employing AdS/CFT correspondence to
get solutions that characterize baryon configuration \cite{10}. More
than these conclusions, the role of cosmic dark energy can be played
by electromagnetic nonlinear fields. In contrast, it is fascinating
to find the exact solutions for Einstein's gravity with a nonlinear
electromagnetic field from the perspective of a black hole (BH)
theory.

In general relativity, one of the most interesting subjects is the
study of BH which is the outcomes of self-gravitating astronomical
objects. These compact objects are the completely collapsed
structure of massive stars defined as thermodynamical objects with
strong gravitational effects such that nothing, not even
electromagnetic radiations such as light escape from them. Black
hole solutions with nonlinear electrodynamics can contribute to
understanding the scientific importance of nonlinear processes in
strong electromagnetic fields and their gravitational impacts. In
this regard, several solutions have been developed for charged BHs,
black branes \cite{11}-\cite{20}, magnetic branes, and magnetic BHs
\cite{21,22,23}. In the literature, apart from general relativity,
solutions in higher derivative gravity with nonlinear
electromagnetic fields have also been investigated
\cite{24}-\cite{26}. By incorporating a nonlinear electromagnetic
field, it is observed that not only the Big-Bang singularity but
also the BH singularity may be prevented. As a result, several
regular BH solutions in the absence of central singularities are
found \cite{27}-\cite{32}. In the presence of nonlinear
electrodynamics, the horizons of BHs are also greatly affected
\cite{33}. Recently, Yu and Gao \cite{34} developed charged BH
solution with nonlinear electrodynamics and found their
singularities are different from Reissner-Nordstr\"{o}m (RN) BH.

The study of wormhole (WH), which is the non-singular solution of
the field equations representing a theoretical framework of geometry
that links two faraway universes or the multi-universes through the
tunnel, is a very interesting topic in astrophysics and cosmology. A
non-traversable WH does not allow two-way observer movement. Due to
the lack of event horizon as well as singularity, the only approach
to travel among distant or multi-universes is through a traversable
WH \cite{34a}. Among these structures, the observer is not able to
travel due to the rapidly expanding and collapsing nature of the
traversable WH throat. The existence of exotic matter is the primary
component that maintains the WH throat in a stable position which
allows the observer motion. The violation of null and weak energy
conditions shows the presence of matter contents with exotic
properties. There are many researches that discussed WH structure in
different modified theories and their physical chracteristics
\cite{344a}-\cite{344k}.

The amount of exotic matter in WH throat can be minimized by
considering the suitable geometry of WH proposed by Visser \cite{35}
in 1989. He also proposed cut and paste technique to formulate a
traversable WH by matching of two equivalent copies of Schwarzschild
BHs at hypersurface which reduces the number of exotic matter
\cite{36}. The respective matter contents like energy density and
pressure are evaluated through Israel thin-shell formalism
\cite{37}. The developed structure is physically acceptable and its
stability has become a new challenge for researchers in the last
three decades. The linear stable configuration is investigated in
\cite{38} and the effects of charge observed in \cite{39}. This work
is extended by Eiroa and Romero \cite{40} for different BH
geometries and observed the effects of charge as well as
cosmological constant on the stability of WH geometries filled with
Chaplygin gas. Many researchers use different approaches to discuss
the stability of WH through radial perturbation with different
equations of state (EoS) \cite{41}-\cite{55}.

The stable structure of thin-shell WHs developed from Schwarzschild
BH with variable EoS is investigated by Varela \cite{56} in 2015. It
is interesting to mention that these EoS depends on the radius of WH
throat and a real constant $n$. In the presence of variable EoS, the
possible existence of a stable structure is enhanced that depends on
the ranges of $n$ and other physical parameters. The stability of
charged thin-shell WH filled with variable EoS developed from RN BH
is analyzed by Eid \cite{57}. Sharif and Javed \cite{58} extended
these concepts for regular BHs and found that variable EoS greatly
affects the stability of thin-shell WHs. Recently, they also
analyzed the comparison between the stability of thin-shell WH (two
equivalent copies of RN BHs) and thin-shell (inter flat and outer RN
BH) through radial perturbation with variable EoS \cite{59}. It is
found that thin-shell WH is less stable than thin-shell for these
matter distributions. Recently, stability of thin-shell around WH
geometries are discussed in different modified theories of gravity
\cite{59a}-\cite{59c}.

The above discussion indicates that the stability of thin-shell WHs
greatly depends on the choice of BH as well as the EoS. In the
present manuscript, we are interested to explore the stable
configuration of WH geometry developed from RN BH with nonlinear
electrodynamics filled with barotropic and two-variable EoS. The
paper is outlined as follows. Section \textbf{2} explains the
general description of RN BH with nonlinear electrodynamics. Section
\textbf{3} provides a complete discussion of the construction of
thin-shell WH from two equivalent copies of considered BHs through
the cut and paste approach. Section \textbf{4} is devoted to
exploring the stability of developed structures by using radial
perturbation with barotropic, generalized phantomlike, and Chaplygin
variable EoS. In the last section, we conclude all the results.

\section{Black Hole with Nonlinear
Electrodynamic}

The action that represents the minimal coupling of nonlinear
electrodynamic to gravity is given as \cite{34}
\begin{equation}\label{1}
S=\int\left[Y[\psi]+R\right]\frac{\sqrt{-g}}{16\pi}d^4x,
\end{equation}
where
\begin{equation}\nonumber
F_{\gamma\beta}=\nabla_\gamma A_\beta-\nabla_\beta A_\gamma, \quad
\psi=F_{\gamma\beta}F^{\gamma\beta}, \quad \gamma,\beta=0,1,2,3,
\end{equation}
here $A_\gamma$ represents the Maxwell field, $R$ denotes Ricci
scalar and $Y[\psi]$ is a function of $\psi$. The respective field
equations by varying the above action are turn out to be \cite{34}
\begin{equation}\label{2}
G_{\gamma\beta}=\frac{1}{2}g_{\gamma\beta}Y[\psi]-2Y[\psi],_\psi
F_{\gamma\mu}F^{\mu}_\beta,
\end{equation}
where $Y[\psi],_\psi=\frac{dY[\psi]}{d\psi}$. The corresponding
generalized Maxwell equations are given as
\begin{equation}\label{3}
\nabla_\gamma[Y[\psi],_\psi F^{\gamma\beta}]=0.
\end{equation}
The spherical symmetric static spacetime is parameterized as
\begin{equation}\label{4}
ds^2=-\Phi(r)dt^2+\Phi^{-1}(r)dr^2 +r^2d\theta^2+r^2\sin^2\theta
d\phi^2.
\end{equation}
here $\Phi(r)$ is the metric function of the spacetime. It is noted
that the non-zero component of Maxwell field tensor for such
spacetime is $A_0=\phi(r)$ and $\psi=-2\phi'^2$. For specific
expression of $Y[\psi]=-2\alpha\sqrt{-2\psi}+\psi$ where $\alpha$ is
coupling constant, the solution of field equations given as
\cite{34}
\begin{eqnarray}\label{5}
\phi (r)&=&-q/r-r\alpha ,\\\label{6} \Phi(r)&=&-\frac{2 m}{r}+2
\alpha q+\frac{q^2}{r^2}-\frac{\alpha ^2 r^2}{3}+1,
\end{eqnarray}
where $q$ is the charge and $m$ denotes the mass of BH. It is
interesting to mention that this solution is reduced to RN BH for
$\alpha=0$ and Schwarzschild BH is recovered when both $\alpha$ and
$q$ are vanished.

In the following, we are interested to develop thin-shell WHs by
considering two equivalent copies of RN BH with nonlinear
electrodynamics. We consider the cut and paste approach to develop
thin-shell WHs.

\section{Formalism of Thin-Shell Wormholes}

As we know that the geometrical structure of WHs connects two
different as well as distant regions of the spacetimes through a
tunnel named WH throat. The study of observers traveling from one
region to another by using WH throat is an interesting topic in
cosmology and astrophysics. The rapidly expanding and collapsing
phenomena of the WH throat do not allow any observe to move freely
through the WH throat. For traversable WH, a specific type of matter
distribution must be required to overcome the collapsing behavior of
the WH throat. The normal matter distribution is not suitable for
the traversable WH, so, there must exist some matter distribution
with exotic properties. Such type of matter does not obey null and
weak energy conditions and is also named exotic matter. To reduce
the amount of exotic matter, Visser introduced the cut and paste
technique to construct thin-shell WHs by joining two equivalent
copies of BH spacetimes at hypersurface. In the present manuscript,
we use this approach to develop the geometry of thin-shell WHs in
the background of two equivalent copies of BH with nonlinear
electrodynamic effects. For this purpose, we cut this spacetime into
the following regions as
\begin{eqnarray}\label{7}
\mathcal{M}^{\pm}=\left\lbrace r^{\pm}\leq
\Re,\Re>r_{h}\right\rbrace,
\end{eqnarray}
where $\Re$ is known as WH throat radius and $r_h$ represents the
radius of the event horizon. These manifolds are connected at
(2+1)-dimensional manifold referred as hypersurface given as
\begin{eqnarray}\label{8}
\Sigma=\left\lbrace r^{\pm}=\Re,\Re>r_{h}\right\rbrace.
\end{eqnarray}

This procedure gives a unique regular manifold and mathematically it
can be expressed as $\mathcal{M}=\mathcal{M}^{-}\cup
\mathcal{M}^{+}$. It is interesting to mention that the event
horizon and singularity in the developed structure can be avoided by
using $\Re>r_h$. According to the Darmois-Israel formalism, the
coordinates of considered manifolds and hypersurface are in the
following form $x^{\gamma}=(t,r,\theta,\phi)$ and
$\xi^{i}=(\tau,\theta,\phi)$, respectively. Here $\tau$ represents
the proper time over the hypersurface. These coordinate systems are
related to one another by using the following coordinate
transformation
\begin{eqnarray}\label{9}
g_{ij}=\frac{\partial x^{\gamma}}{\partial\xi^{i}}\frac{\partial
x^{\beta}}{\partial\xi^{j}}g_{\gamma\beta}.
\end{eqnarray}
The respective parametric equation for the hypersurface is defined
as
\begin{eqnarray}\nonumber
\Sigma:R(r,\tau)=r-\Re(\tau)=0.
\end{eqnarray}
The dynamical configuration of a thin shell is studied by
considering the dependence of shell radius ($\Re$) over the proper
time on the shell. Therefore, the shell radius can be expressed as a
function of proper time as $\Re=\Re(\tau)$. The corresponding
induced metric has the following form
\begin{equation}\label{10}
ds_{\Sigma}^{2}=\Re^2(\tau)\sin^2\theta
d\phi^2+\Re^2(\tau)d\theta^2-d\tau^2.
\end{equation}
The matter contents of the shell have remarkable importance over the
stability and dynamics of the WH throat. The physical quantities of
matter distribution are evaluated through the reduced form of
Einstein field equations at hypersurface. These equations are also
known as Lanczos equations given as
\begin{eqnarray}\label{11}
{S^{i}}_{j}=-\frac{1}{8\pi}\left(\left[{K^{i}}_{j}\right]
-{\delta^{i}}_{j}\,K\right).
\end{eqnarray}
where ${K^{i}}_{j}$ represents the components of extrinsic
curvature, $K$ is the trace of extrinsic curvature
($K=[{K^{i}}_{i}]$) and
${S^{i}}_{j}=diag(\rho,\mathcal{P},\mathcal{P})$ denotes
stress-energy tensor. The surface energy density and pressure of the
matter distribution located at $\Sigma$ is denoted with $\rho$ and
$\mathcal{P}$, respectively. Such matter distribution produces
discontinuity in the inner and outer components of extrinsic
curvature which mathematically given as
$[{K^{i}}_{j}]={K^{+i}}_{j}-{K^{-i}}_{j}\neq0$. The extrinsic
curvature of interior and exterior geometries are defined as
\begin{eqnarray}\label{12}
K_{ij}^{(\pm)}=-n_{\mu}^{(\pm)}\left(\frac{\partial^{2}x^{\mu}}
{\partial\xi^{i}\partial\xi^{j}}+\Gamma_{\gamma\beta}^{\mu}\frac{\partial
x^{\gamma}}{\partial\xi^{i}}\frac{\partial
x^{\beta}}{\partial\xi^{j}}\right)_{\Sigma}.
\end{eqnarray}

The temporal and radial components of unit normals over
$\mathcal{M}^{\pm}$ become
\begin{eqnarray}\nonumber
n^{\pm}_{t}&=&-\dot{\Re},\\\nonumber
n^{\pm}_{r}&=&\frac{\sqrt{\dot{\Re}^2+\frac{q^2-2 m \Re}{\Re^2}+2
\alpha q-\frac{1}{3} \alpha ^2 \Re^2+1}}{\frac{q^2-2 m \Re}{\Re^2}+2
\alpha q-\frac{1}{3} \alpha ^2 \Re^2+1},
\end{eqnarray}
respectively. The derivative with respect to proper time is denoted
with overdot. The corresponding extrinsic curvature components are
given as
\begin{eqnarray}\label{13}
K_{\tau\tau}^{\pm}&=&\pm\frac{\frac{2 m}{\Re^2}-\frac{2
q^2}{\Re^3}-\frac{2 \alpha ^2
\Re}{3}+2\ddot{\Re}}{2\sqrt{\frac{q^2-2 m \Re}{\Re^2}+2 \alpha
q-\frac{1}{3} \alpha ^2 \Re^2+1+\dot{\Re}^2}},\\\label{14}
K_{\theta\theta}^\pm&=&\pm \Re\sqrt{\frac{q^2-2 m \Re}{\Re^2}+2
\alpha q-\frac{1}{3} \alpha ^2 \Re^2+1+\dot{\Re}^2},\\\label{15}
K^{\pm}_{\phi\phi}&=&\sin^2\theta K_{\theta\theta}^\pm.
\end{eqnarray}
By using Eqs.(\ref{13})-(\ref{15}) in Lanczos equations (\ref{11}),
we have
\begin{eqnarray}\label{16}
2\pi\Re\rho&=&-\sqrt{\dot{\Re}^2-\frac{2 m}{\Re}+\frac{q^2}{\Re^2}+2
\alpha  q-\frac{\alpha ^2 \Re^2}{3}+1},
\\\label{17}
8\pi\Re \mathcal{P}&=&\frac{2\dot{\Re}^2+2\Re\ddot{\Re} -\frac{2
m}{\Re}+4 \alpha q-\frac{1}{3} 4 \alpha ^2
\Re^2+2}{\sqrt{\dot{\Re}^2+-\frac{2 m}{\Re}+\frac{q^2}{\Re^2}+2
\alpha q-\frac{\alpha ^2 \Re^2}{3}+1}}.
\end{eqnarray}

Now, it is assumed that the thin shell of the developed geometry
does not move along its radial direction at equilibrium shell radius
$\Re_0$. Therefore, it is interesting to mention that the proper the
time derivative of shell radius vanishes, i.e.,
$\dot{\Re_0}=0=\ddot{\Re_0}$. Hence, we have
\begin{eqnarray}\label{18}
4 \pi  \Re_0\rho_{0}&=&-\sqrt{\frac{q^2-2 m \Re_0}{\Re_0^2}+2 \alpha
q-\frac{1}{3} \alpha ^2 \Re_0^2+1},\\\label{19} 4 \pi \Re_0^2
\mathcal{P}_{0}&=&\frac{\Re_0 \left(6 \alpha  q-2 \alpha ^2
\Re_0^2+3\right)-3 m}{ \sqrt{\frac{9 \left(q^2-2 m
\Re_0\right)}{\Re_0^2}+18 \alpha q-3 \alpha ^2 \Re_0^2+9}},
\end{eqnarray}
where surface energy density and pressure at the equilibrium
position are denoted with $\rho_0$ and $\mathcal{P}_0$,
respectively. There are three well-known energy conditions, i.e.,
weak ($\mathcal{P}+\rho\geq0$, $\rho\geq0$), null
($\mathcal{P}+\rho\geq0$) and strong ($3\mathcal{P}+\rho\geq0$)
energy conditions. Here, we analyze the energy conditions
graphically for different values of $\alpha$ as shown in Fig. 1. It
is noted the surface energy density is negative ($\rho_{0}<0$) which
leads to the violation of weak as well as the dominant energy
constraints. Such violations indicate that the developed structure
is filled with matter distribution having exotic nature. These
matter distributions at the WH throat produce repulsion against
collapse and also helps to keep it open. Hence, the developed the
structure is physically acceptable for the wormhole configuration.
We conclude that the presence of surface matter in WH throat
violates the null, weak, and strong energy conditions as shown in
Fig. 1.

By considering the values of energy density of the shell (\ref{18}),
we obtain the respective equation of motion given as
\begin{eqnarray}
\dot{\Re}^{2}=-\Omega(\Re),\label{20}
\end{eqnarray}
the effective potential of the shell is defined as
\begin{equation}\label{21}
\Omega(\Re)= 2 \alpha q-4 \pi ^2  \Re^2 \rho^2 -\frac{2 m}{
\Re}+\frac{q^2}{ \Re^2}-\frac{\alpha ^2  \Re^2}{3}+1.
\end{equation}

The repulsive and attractive characteristics of WH throat is
evaluated through the 4-acceleration of the observer as
\begin{equation}\nonumber
a^\beta=v^\beta_{;\alpha}v^\alpha,
\end{equation}
here 4-velocity of an observer is denoted with
$v^\alpha=\left(\frac{1}{\sqrt{\Phi}},0,0,0\right)$. The respective
equation of motion turns out to be
\begin{equation}\nonumber
\frac{d^2r}{d\tau^2}=-\Gamma^{r}_{tt}\left(\frac{dt}{d\tau}\right)^2=-a^r,
\end{equation}
which yields
\begin{equation}\nonumber
a^r=\frac{m}{\Re^2}-\frac{q^2}{\Re^3}-\frac{\alpha ^2 \Re}{3}.
\end{equation}
\begin{figure}\centering
\epsfig{file=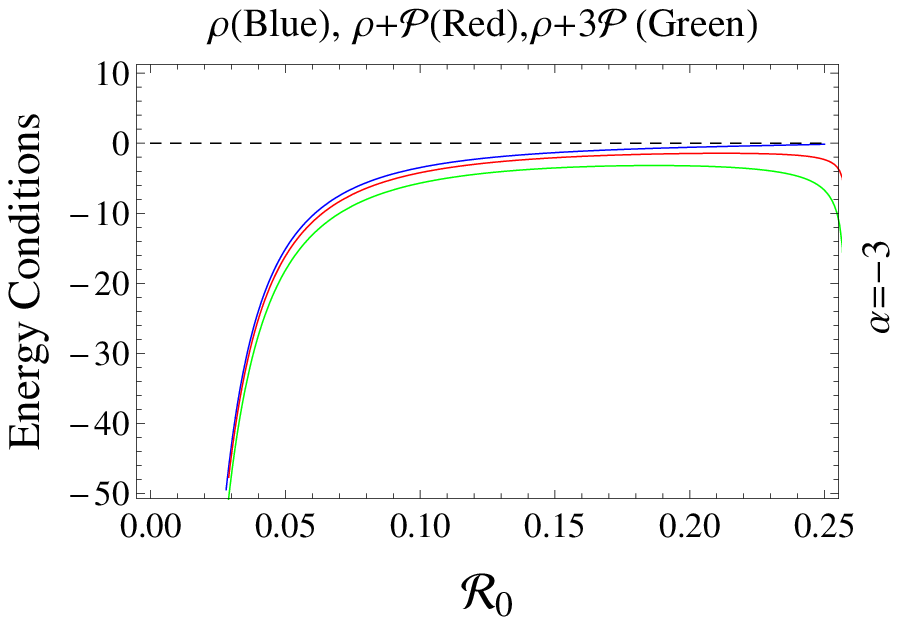,width=.45\linewidth}\epsfig{file=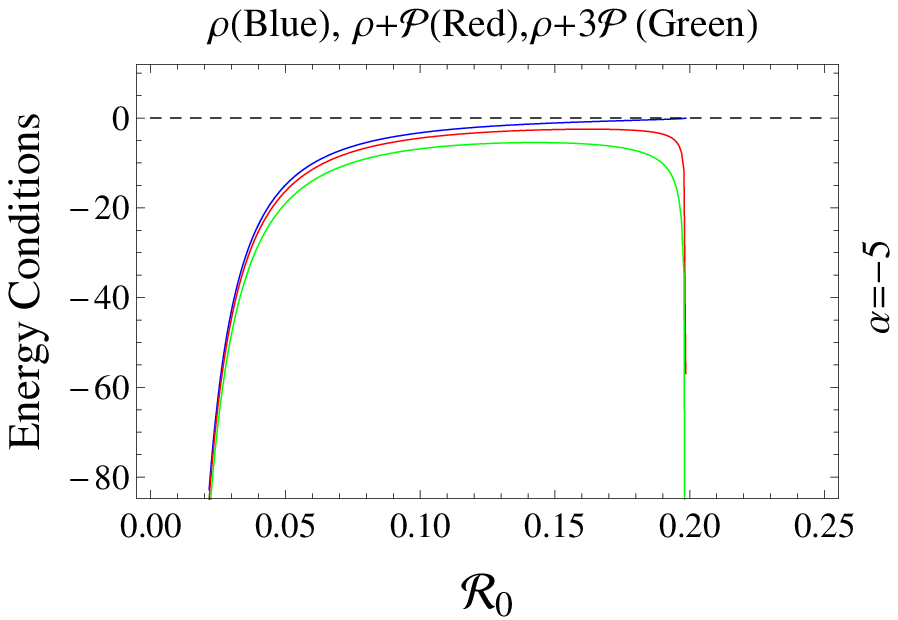,width=.45\linewidth}
\caption{Energy conditions of thin-shell WHs for different values of
$\alpha$.}
\epsfig{file=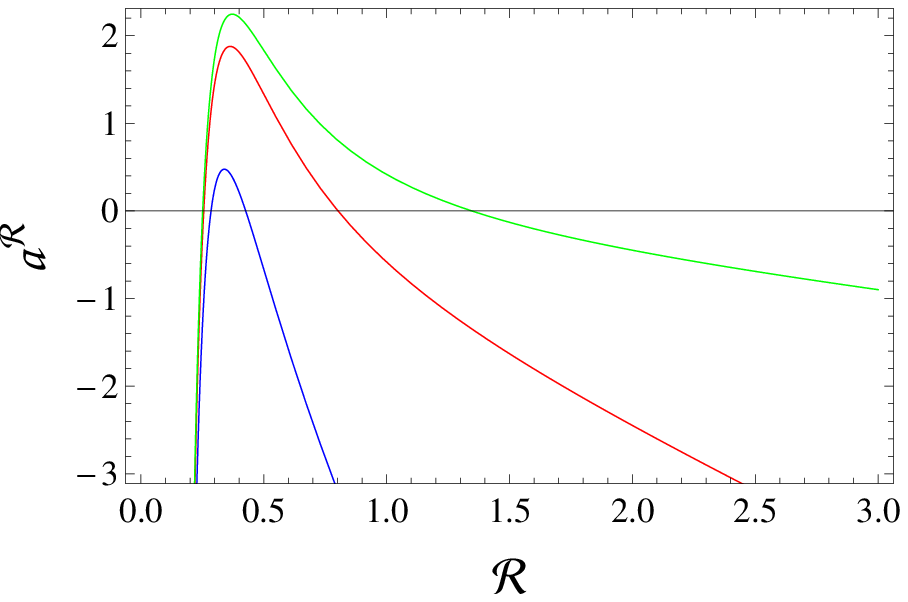,width=.45\linewidth}\epsfig{file=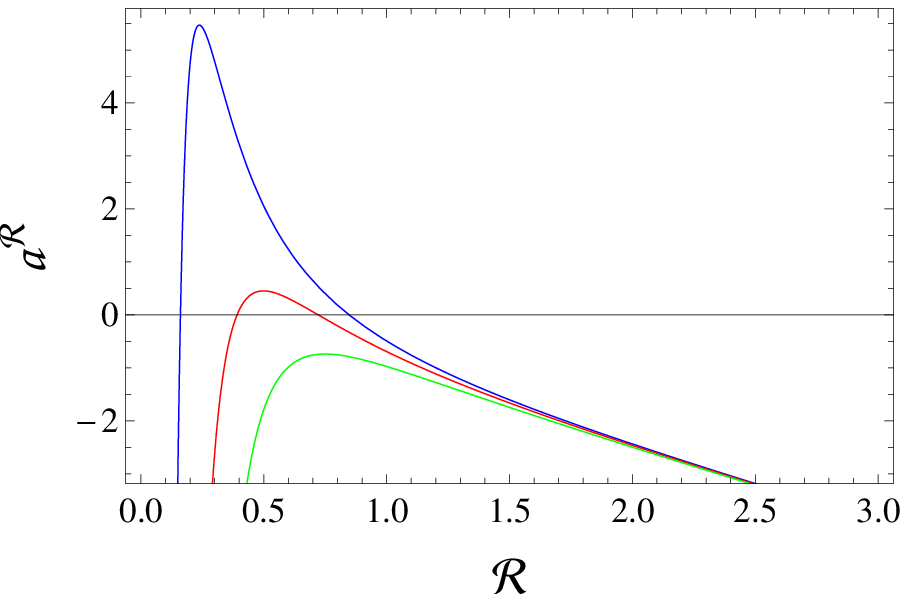,width=.45\linewidth}
\caption{ Thin-shell WHs repulsive and attractive characteristics
for different values of $\alpha$ as $\alpha=-4$(blue),
$\alpha=-2$(red), $\alpha=1$(green) with $m=1$, $q=0.5$ (left plot)
and $q=0.4$(blue), $q=0.6$(red), $q=0.8$(green) with $m=1$,
$\alpha=-2$ (right plot).}
\end{figure}
The 4-acceleration radial component explains the repulsive ($a^r<0$)
as well as the attractive ($a^r>0$) nature of the throat. It is
found that
\begin{itemize}
\item For repulsive nature, an
inside directed radial acceleration is needed to avoid the effect of
pushed away by the WH.
\item For attractive nature, an outward-directed radial acceleration is
required to overcome the WH attraction.
\end{itemize}
It is noted that the radial acceleration decreases as shell radius
increases as shown in Fig. 2. For highly negative values of coupling
constant $\alpha$, thin-shell shows large repulsive nature as shell
radius increases. We find that the repulsive behavior of the shell
decreases as coupling constant approaches to positive values (left
plot). It is also found that shell indicates the initially
attractive behavior and then repulsive nature for different values
of charge (right plot). The attractive behavior decreases as the
charge of the geometry is enhanced.

\section{Stability Analysis}

For stability analysis, we consider static shell radius $\Re_0$ and
expand the effective potential $\Omega(\Re)$ about $\Re_0$ by using
Taylor series upto second order terms as follows
\begin{equation}\label{22}
\Omega(\Re)=\Omega(\Re_{0})+\Omega'(\Re_{0})(\Re-\Re_{0})+\frac{1}{2}
\Omega''(\Re_{0})(\Re-\Re_{0})^2+O[(\Re-\Re_{0})^3].
\end{equation}
It is interesting to mention that the stable and unstable geometry
of the wormhole, the throat requires that the potential function and
its the first derivative must be vanished at the equilibrium
position, i.e., $\Omega(\Re_0)=0=\Omega'(\Re_0)$. Then, it can be
evaluated as:
\begin{itemize}
\item If the second derivative of the potential at $\Re=\Re_0$ is positive then
it represents the stable configuration and expressed unstably
structure if $\Omega''(\Re_0)<0$.
\item It is neither stable nor unstable if $\Omega''(\Re_0)=0$.
\end{itemize}
Hence, for equilibrium configuration, Eq.(\ref{22}) becomes
\begin{equation}\label{23}
\Omega(\Re)=\frac{1}{2}\Omega''(\Re_{0})(\Re-\Re_{0})^2.
\end{equation}

It is noted that $\rho$ and $\mathcal{P}$ obey the conservation
equation given as
\begin{equation}\label{24} \mathcal{P} \frac{d}{d\tau}(4\pi
\Re^2)+\frac{d}{d\tau}(4\pi \Re^2\rho)=0.
\end{equation}
The exact solution of the conservation equation depends on the
choice of matter distribution which can be described through EoS.
Here, we consider two types of EoS $\mathcal{P}=\mathcal{P}(\rho)$
and $\mathcal{P}=\mathcal{P}(\rho,\Re)$ \cite{56}. The second case
is more general in which surface pressure of the shell depends on
both surface energy density and the throat radius. For both cases,
we have $\mathcal{P}'=\frac{d\mathcal{P}(\rho)}{d\rho}\rho'$ and
$\mathcal{P}'=\frac{d\mathcal{P}}{d\rho}\rho'+\frac{d\mathcal{P}}{d\Re}$,
respectively. Hence, The conservation equation is yield
\begin{equation}\label{25}
\rho'=-\frac{2}{\Re}\{\rho+\mathcal{P}(\rho,\Re)\}.
\end{equation}
For every choice of variable EoS, each solution of Eq.(\ref{25})
leads to a specific form of $\Omega(\Re)$. The second derivative of
effective potential at $\Re=\Re_{0}$ given as
\begin{eqnarray}\nonumber
\Omega''(\Re_0)&=&\frac{6
q^2}{\Re^4_0}-16\pi^2\left(1+2\beta_{0}^2\right)(\rho^2_0+\rho_0
\mathcal{P}_0)+8\pi^2(\rho_0+2\mathcal{P}_0)^2-\frac{4
m}{\Re^3_0}\\\label{26}&-&\frac{2 \alpha
^2}{3}+16\pi^2\Re_0\rho_0\gamma_{0}.
\end{eqnarray}
where $\gamma_{0}=\frac{d\mathcal{P}}{d\Re}\mid_{\Re=\Re_0}$ and
$\beta_{0}^2=\frac{d\mathcal{P}}{d\rho}\mid_{\Re=\Re_0}$. It is
found that $\Omega''(\Re_{0})$ depends on EoS parameters
$\gamma_{0}$ and $\beta_{0}^2$.

In the following, we observe the effects of barotropic and variable
EoS on the stability of the developed geometry.

\subsection{Barotropic EoS}

In our first case, we choose barotropic model to explore the
stability of thin-shell WHs. It linearly relates the surface
pressure and energy density of the matter contents given as
\begin{equation}\label{27}
\mathcal{P}=\varpi \rho,
\end{equation}
where $\varpi$ denotes the barotropic EoS parameter. By considering
Eq.(\ref{27}) in (\ref{25}), we get
\begin{equation}\label{28}
\rho'(\Re)=-\frac{2}{\Re}(1+\varpi)\rho(\Re),
\end{equation}
which yields
\begin{equation}\label{29}
\rho(\Re)=\rho(\Re_{0})
\left(\frac{\Re_{0}}{\Re}\right)^{2(1+\varpi)}.
\end{equation}
The respective expression of the effective potential turns out to be
\begin{equation}\label{30}
\Omega(\Re)=-4 \pi ^2 \Re^2 \rho _0^2
\left(\frac{\Re_0}{\Re}\right)^{4 (\varpi+1)}-\frac{2
m}{\Re}+\frac{q^2}{\Re^2}+2 \alpha  q-\frac{\alpha ^2 \Re^2}{3}+1,
\end{equation}
the first derivative with respect to throat radius $``\Re"$ at
$\Re_{0}$ becomes
\begin{equation}\nonumber
\Omega'(\Re_{0})=\frac{4 a q^2}{\Re_{0}^3}-\frac{8 a
m}{\Re_{0}^2}+\frac{8 \varpi \alpha q}{\Re_{0}}-\frac{4}{3} \varpi
\alpha ^2 \Re_{0}+\frac{4 \varpi}{\Re_{0}}-\frac{2
m}{\Re_{0}^2}+\frac{4 \alpha q}{\Re_{0}}-\frac{4 \alpha ^2
\Re_{0}}{3}+\frac{2}{\Re_{0}}.
\end{equation}
It is noted that $\Omega'(\Re_{0})$ vanishes if and only if
\begin{equation}\label{31}
\varpi=\frac{\Re_{0} \left(3 m-6 \alpha  q \Re_{0}+2 \alpha ^2
\Re_{0}^3-3 \Re_{0}\right)}{2 \left(-6 m \Re_{0}+3 q^2+6 \alpha  q
\Re_{0}^2-\alpha ^2 \Re_{0}^4+3 \Re_{0}^2\right)}.
\end{equation}
For particular value of $\varpi=-1$, we have obtain the position of
equilibrium shell radius as
\begin{equation}\label{32}
\Re_{0}=\frac{\sqrt{9 m^2-16 \alpha  q^3-8 q^2}+3 m}{2 (2 \alpha
q+1)}.
\end{equation}
Also, it is found that
\begin{eqnarray}\nonumber
\Omega''(\Re_{0})&=&\frac{16 \alpha ^2 \varpi^2}{3}+\frac{32
\varpi^2 m}{\Re_{0}^3}-\frac{16 \varpi^2 q^2}{\Re_{0}^4}-\frac{32
\alpha \varpi^2 q}{\Re_{0}^2}-\frac{16 \varpi^2}{\Re_{0}^2}+\frac{4
\alpha ^2}{3}+\frac{20 \alpha ^2 \varpi}{3}\\\label{33}&+&\frac{40
\varpi m}{\Re_{0}^3}-\frac{20 \varpi q^2}{\Re_{0}^4}-\frac{40 \alpha
\varpi q}{\Re_{0}^2}-\frac{20 \varpi}{\Re_{0}^2}+\frac{8
m}{\Re_{0}^3}-\frac{12 \alpha q}{\Re_{0}^2}-\frac{6}{\Re_{0}^2}.
\end{eqnarray}
The above equation is very useful to explore the stable
characteristics of developed structure. For this purpose, it is
plotted and observed different regions at equilibrium shell radius
as shown in Figures \textbf{2-4}. Here, it is interesting to mention
that different regions mention different characteristics of the
developed structure given as:
\begin{itemize}
\item Blue region shows
neither stable nor unstable structure ($\Omega''(\Re_{0})=0$).
\item Brown
region represents stable geometry ($\Omega''(\Re_{0})>0$).
\item Magenta region denotes unstable configuration
($\Omega''(\Re_{0})<0$).
\end{itemize}

It is noted that the developed structure expresses stable behavior
if $\alpha<0$ and $q>0.5$ otherwise it represents unstable or
neither stable nor unstable behavior (Fig. 3). For positive values
of $\alpha$, thin-shell shows the only unstable configuration for
every choice of other physical parameters. The stability of WH is
increased for higher values of charged and decreased for smaller
values. Similarly, the stable structure is explored along with mass,
charge as well as $\alpha$ and found similar results (Figs. 3 and
5). The maximum stable regions are formed for highly negative values
of $\alpha$ and higher values of charge (Fig. 5). It is found that
the presence of nonlinear electrodynamics gives the possibility of
stable configuration in the background of barotropic type fluid
distribution. Furthermore, we plotted the second derivative of
effective potential at equilibrium shell radius for these mentioned
stable regions as shown in Fig. 6. It is noted that thin-shell shows
stable behavior for different values of mass with negative values of
$\alpha$ and stability decreases for higher values of mass. We found
that the stability of WHs greatly affected by the coupling parameter
and its positive as well as negative values. For higher negative
values of $\alpha$, we have obtained a more stable structure as
compared to less negative or positive values (right plot of Fig. 6).
\begin{figure}\centering
\epsfig{file=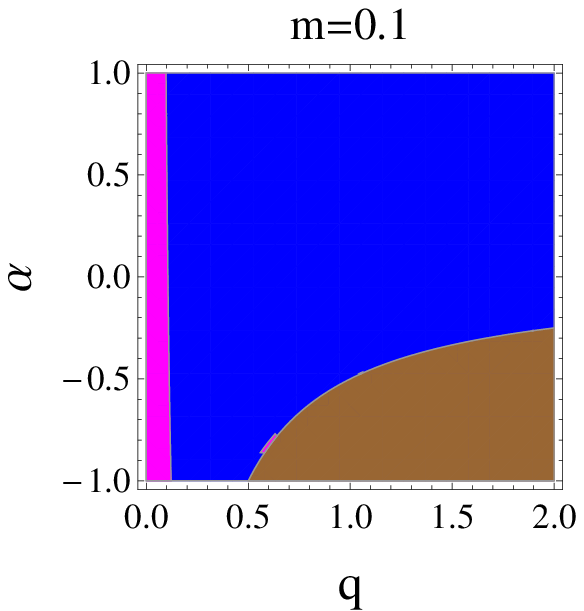,width=.325\linewidth}\epsfig{file=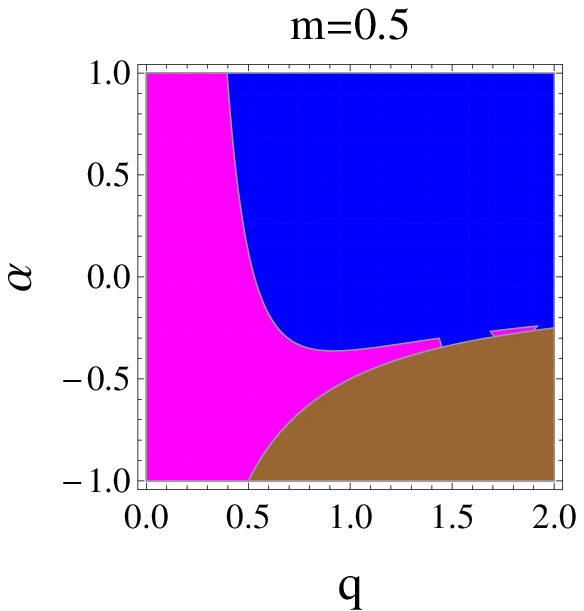,width=.325\linewidth}\epsfig{file=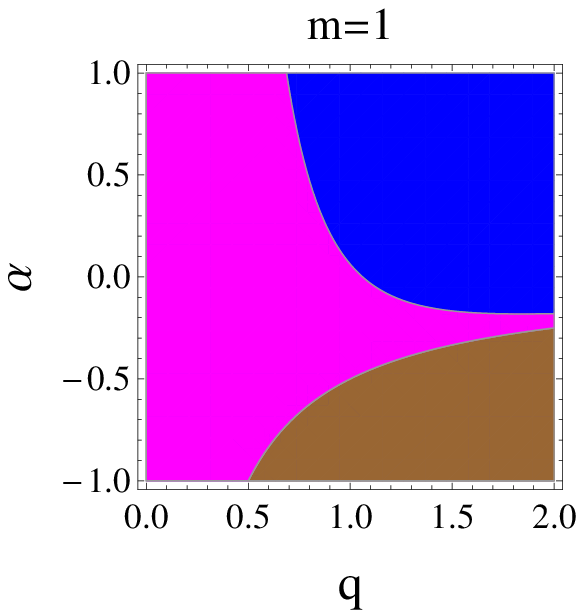,width=.325\linewidth}
\caption{Region plots of $\Omega''(\Re_{0})$ verses $q$ and $\alpha$
for barotropic EoS with different values of $m$.}
\end{figure}
\begin{figure}\centering
\epsfig{file=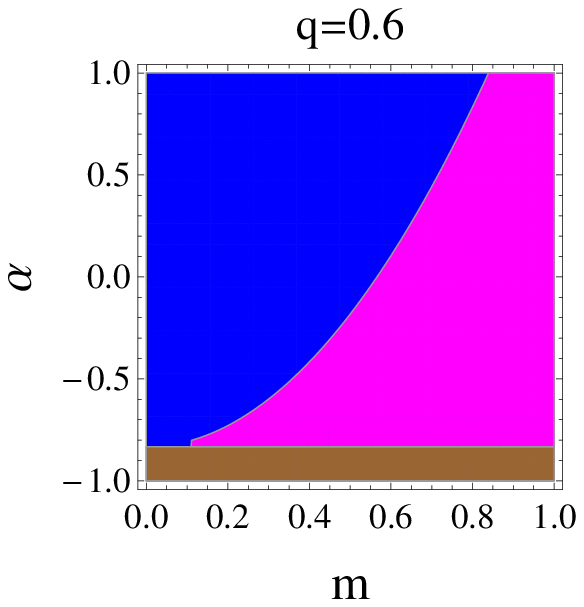,width=.325\linewidth}\epsfig{file=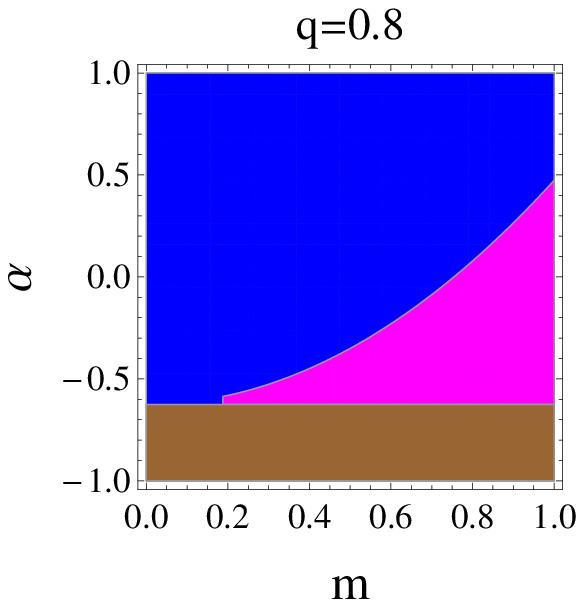,width=.325\linewidth}\epsfig{file=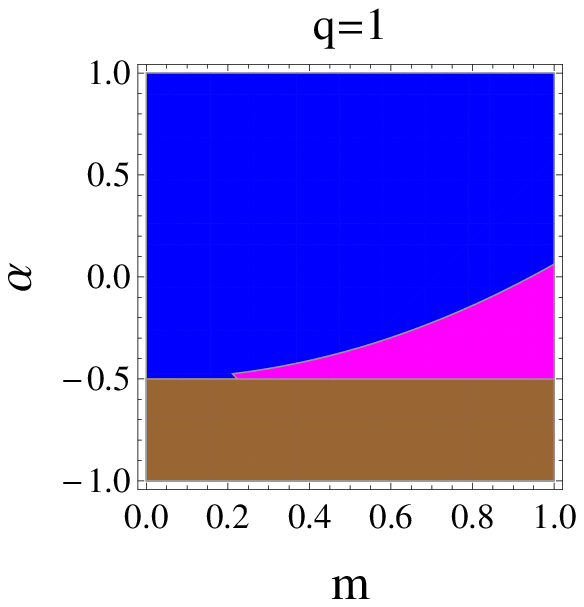,width=.325\linewidth}
\caption{Region plots of $\Omega''(\Re_{0})$ verses $m$ and $\alpha$
for barotropic EoS  with different values of $q$.}
\end{figure}
\begin{figure}\centering
\epsfig{file=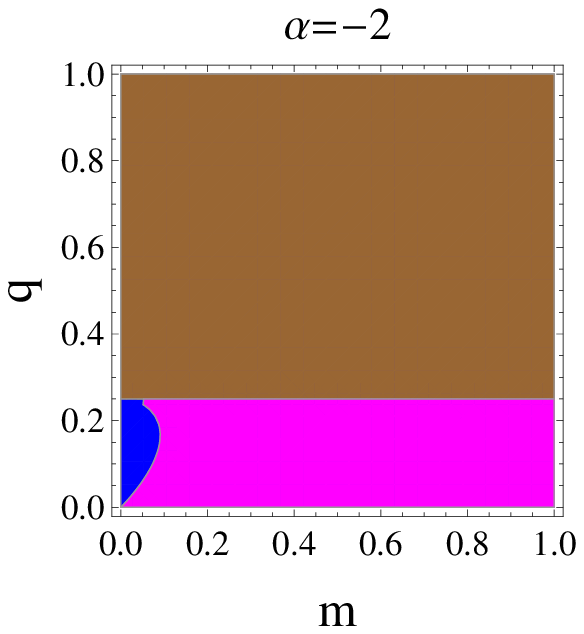,width=.325\linewidth}\epsfig{file=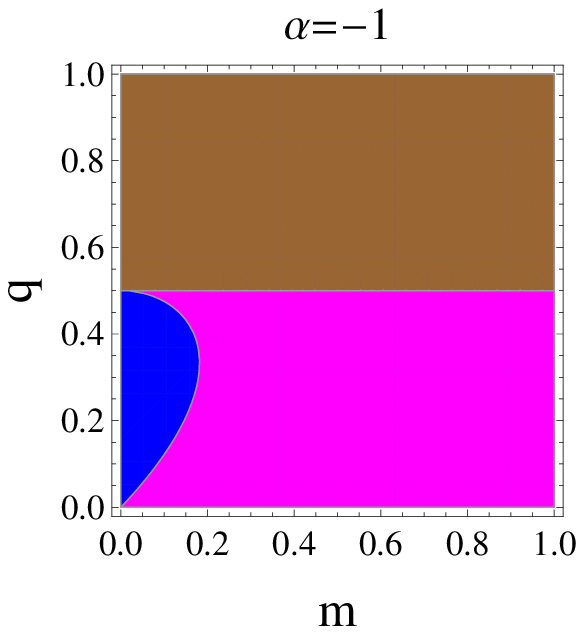,width=.325\linewidth}\epsfig{file=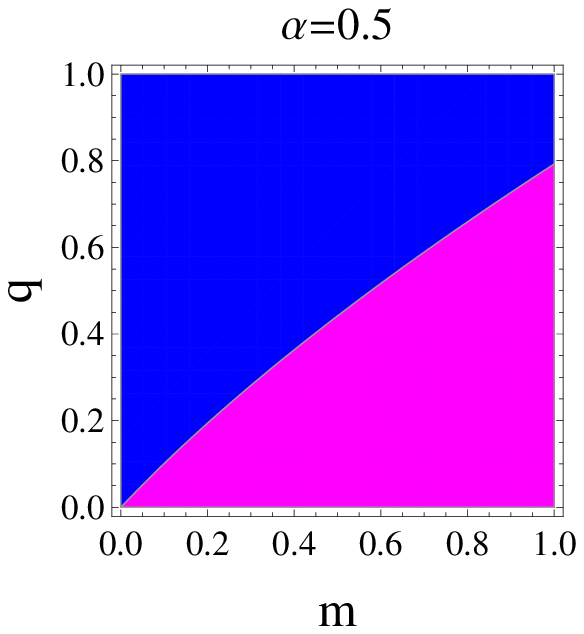,width=.325\linewidth}
\caption{Region plots of $\Omega''(\Re_{0})$ verses $m$ and $q$ for
barotropic EoS with different values of $\alpha$.}
\end{figure}
\begin{figure}\centering
\epsfig{file=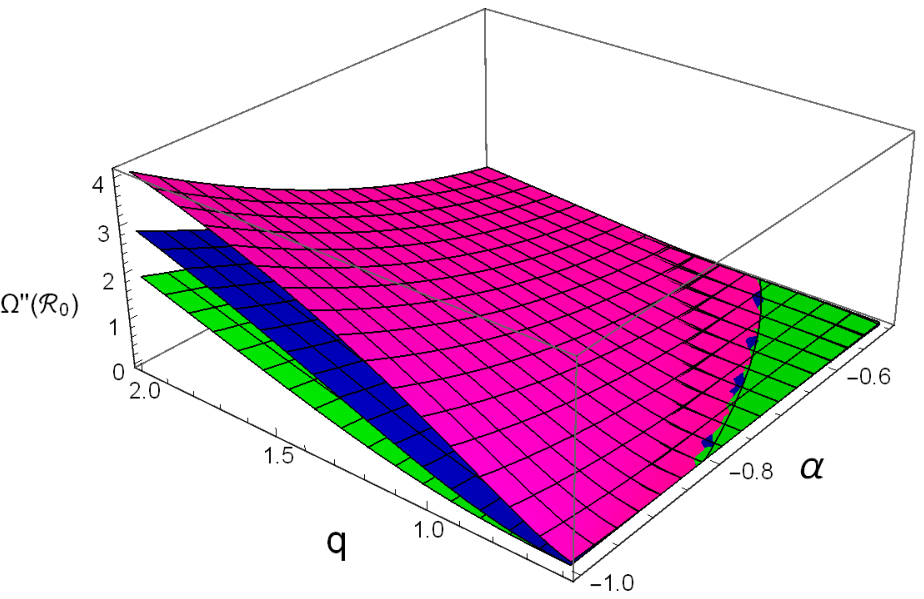,width=.5\linewidth}\epsfig{file=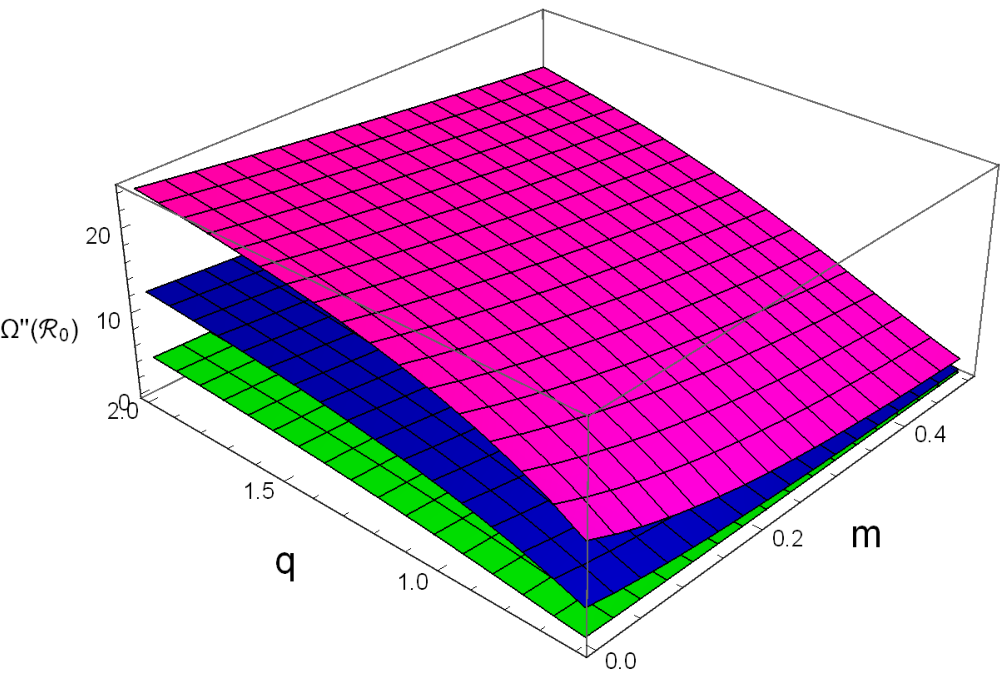,width=.5\linewidth}
\caption{Stability of thin-shell WH for different values of mass as
$m=0.1$(magenta) $m=0.5$(blue) $m=1$(green) (left plot) and $\alpha$
as $\alpha=-2$(magenta) $\alpha=-1.5$(blue) $\alpha=-1$(green)
(right plot) in the background of barotropic type fluid
distribution.}
\end{figure}

\subsection{Phantomlike Variable EoS}

For second case, we consider phantomlike variable EoS to discuss the
stability of thin-shell WHs \cite{56}. Mathematically, it can be
expressed as
\begin{equation}\label{34}
\mathcal{P}=\frac{W}{\Re^n}\rho,
\end{equation}
where $n$ is a real constant and $W$ is the EoS parameter. This
equation is the generalized form of phantomlike EoS. It is reduced
to phantomlike EoS for $n \rightarrow 0$. By using this EoS, the
solution of conservation equation yields
\begin{equation}\label{35}
\rho(\Re)=\rho_{0}e^{\frac{ W}{n}
\left(\frac{1}{\Re^n}-\frac{1}{\Re_0^n}\right)}\left(\frac{\Re_{0}}{\Re}\right)^2.
\end{equation}
The corresponding effective potential turns out to be
\begin{equation}\label{36}
\Omega(\Re)=-4 \pi ^2 \Re^2 \rho _0^2
\left(\frac{\Re_0}{\Re}\right)^4 e^{\frac{2 W}{n}
\left(\frac{1}{\Re^n}-\frac{1}{\Re_0^n}\right)}-\frac{2
m}{\Re}+\frac{q^2}{\Re^2}+2 \alpha  q-\frac{\alpha ^2 \Re^2}{3}+1.
\end{equation}
It is noted that effective potential vanishes at $\Re=\Re_0$ and its
first derivative is given as
\begin{equation}\label{37}
\Omega'(\Re_0)=\frac{1}{3} \Re_0^{-n-3} \left(24 \pi ^2 \rho _0^2
\Re_0^4 \left(W+\Re_0^n\right)-2 \Re_0^n \left(-3 m \Re_0+3
q^2+\alpha ^2 \Re_0^4\right)\right),
\end{equation}
By taking $\Omega'(\Re_{0})=0$, we obtain
\begin{equation}\label{38}
W=\frac{-3 m \Re_0^{n+1}+6 \alpha  q \Re_0^{n+2}-2 \alpha ^2
\Re_0^{n+4}+3 \Re_0^{n+2}}{-3 \left(\Re_0 (\Re_0-2 m)+q^2\right)-6
\alpha q \Re_0^2+\alpha ^2 \Re_0^4}.
\end{equation}
Hence, we have
\begin{eqnarray}\nonumber
\Omega''(\Re_0)&=&-\frac{2}{3} \Re_0^{-2 (n+2)} \left(\left(2 W^2+W
(n+5) \Re_0^n+3 \Re_0^{2 n}\right) \left(3 q^2+6 \alpha  q
\Re_0^2\right.\right.\\\label{39}&-&\left.\left.\Re_0 \left(6
m+\alpha ^2 \Re_0^3-3 \Re_0\right)\right)+\Re_0^{2 n} \left(6 m
\Re_0-9 q^2+\alpha ^2 \Re_0^4\right)\right).
\end{eqnarray}
The stability of thin-shell filled with phantomlike variable EoS is
analyzed graphically. It is noted that the stable regions exist if
and only if both $n$ and $\alpha$ are positive or negative (Figs.
7-9). If $\alpha<0$, then $n$ must be negative for stability with
every value of physical parameters. Fig. 9 indicates that stable
regions occurs when $\alpha>0$ and $n>5$. Hence, for such type of
matter contents, there must exist a possibility of stable regions at
which $\Omega''(\Re_0)>0$ for specific values of mass and charge.
For large stable regions, it is noted that both $\alpha$ and $n$
must be negative. We also noted that stability of the shell
increases for highly negative values of $n$ and $\alpha$ as shown in
left and right plots of Fig. 10.
\begin{figure}\centering
\epsfig{file=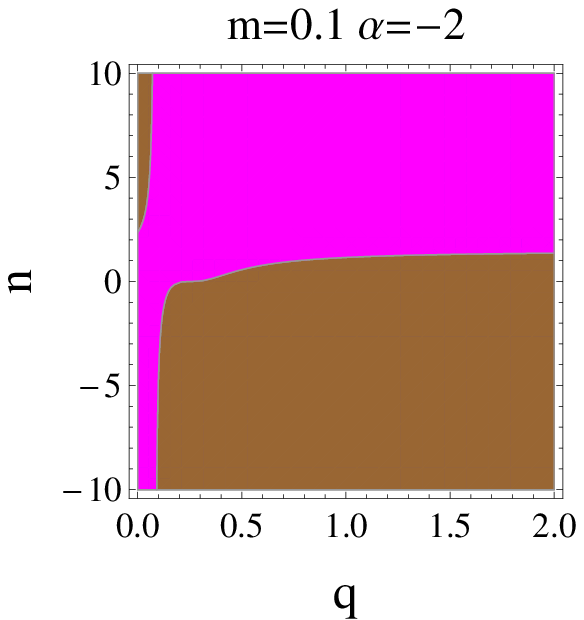,width=.325\linewidth}\epsfig{file=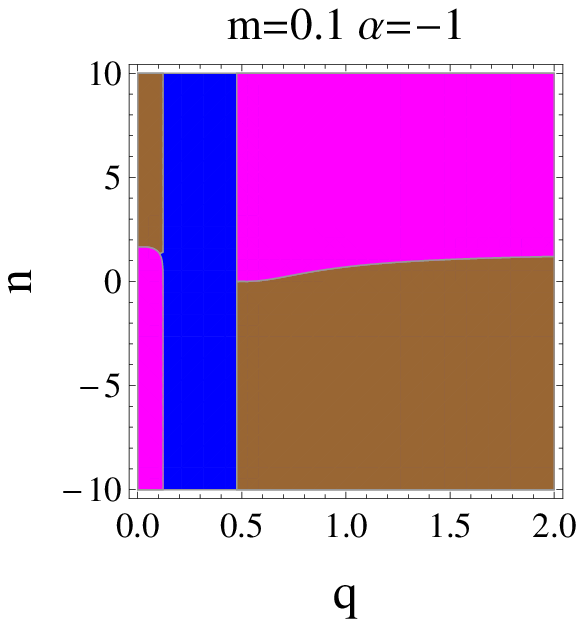,width=.325\linewidth}\epsfig{file=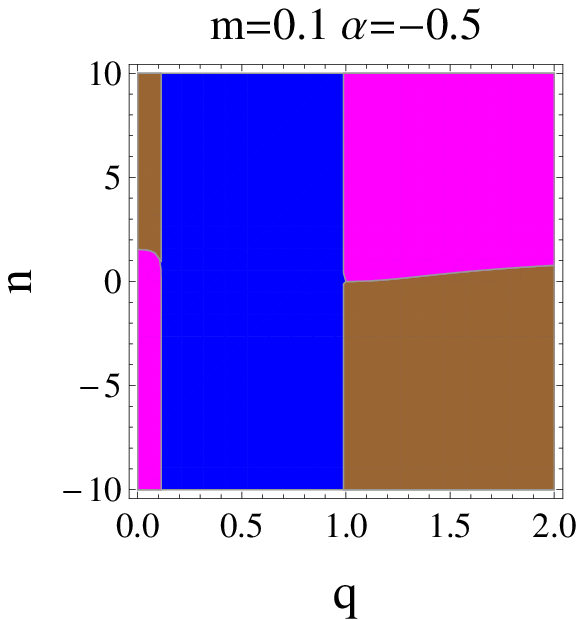,width=.325\linewidth}
\caption{Region plots of $\Omega''(\Re_{0})$ verses $q$ and $n$ for
phantomlike variable EoS with different values of $\alpha$.}
\end{figure}
\begin{figure}\centering
\epsfig{file=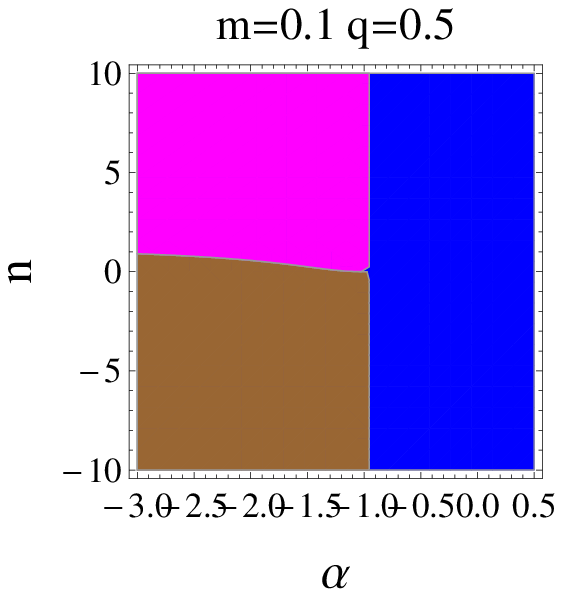,width=.325\linewidth}\epsfig{file=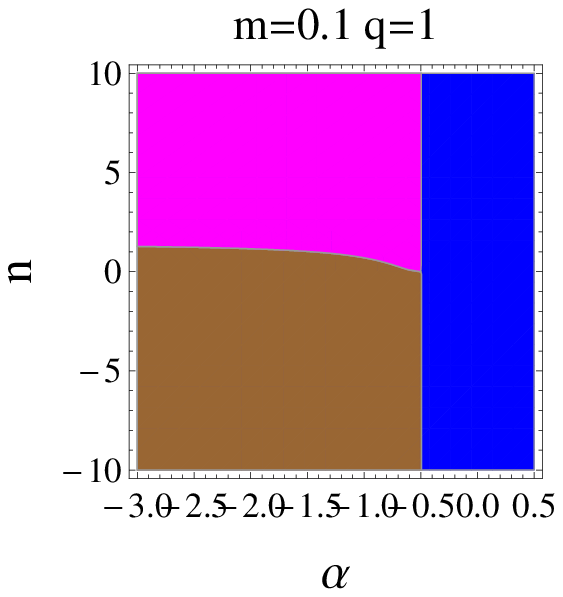,width=.325\linewidth}\epsfig{file=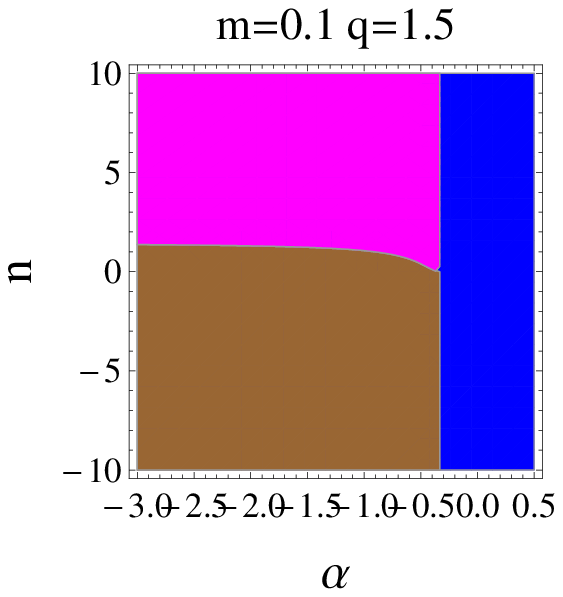,width=.325\linewidth}
\caption{Region plots of $\Omega''(\Re_{0})$ verses $\alpha$ and $n$
for phantomlike variable EoS with different values of $q$.}
\end{figure}
\begin{figure}\centering
\epsfig{file=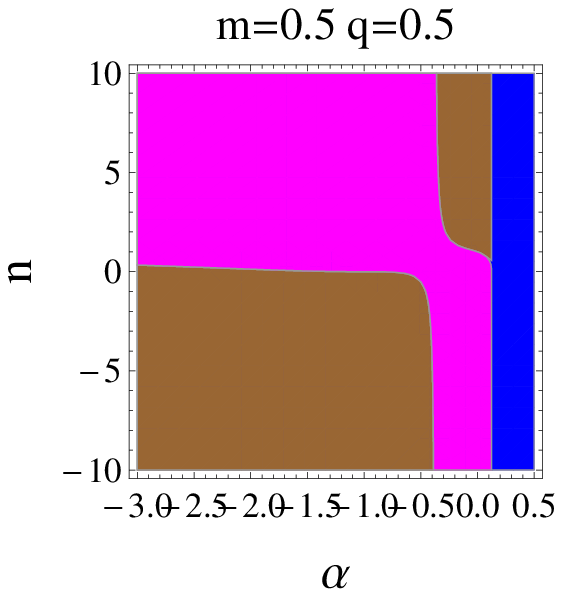,width=.325\linewidth}\epsfig{file=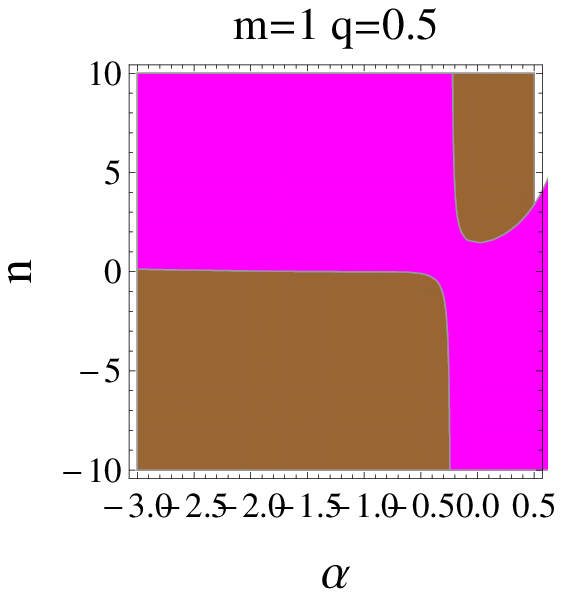,width=.325\linewidth}\epsfig{file=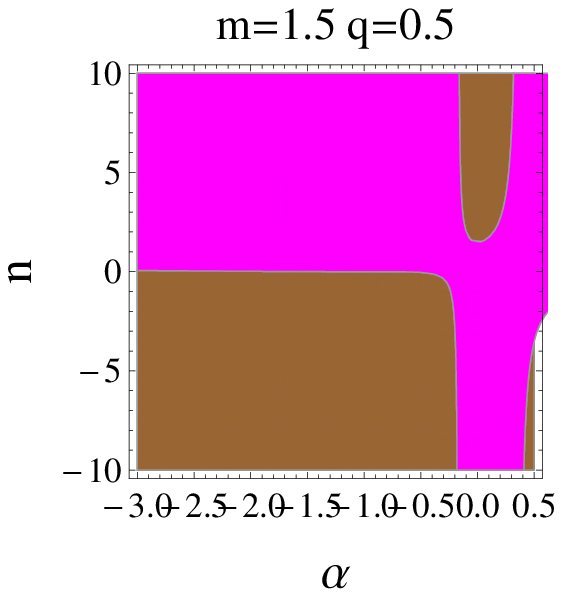,width=.325\linewidth}
\caption{Region plots of $\Omega''(\Re_{0})$ verses $\alpha$ and $n$
for phantomlike variable EoS with different values of $m$.}
\end{figure}
\begin{figure}\centering
\epsfig{file=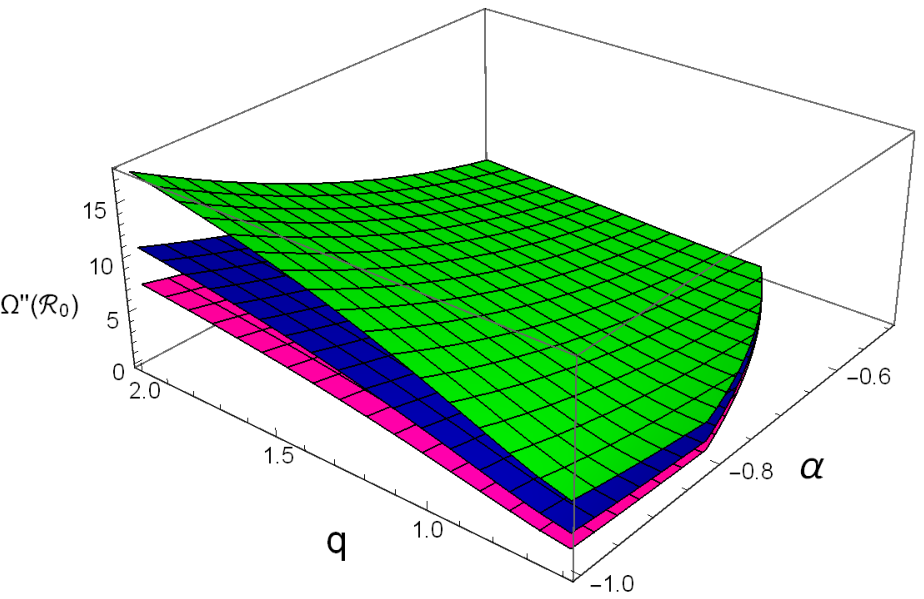,width=.5\linewidth}\epsfig{file=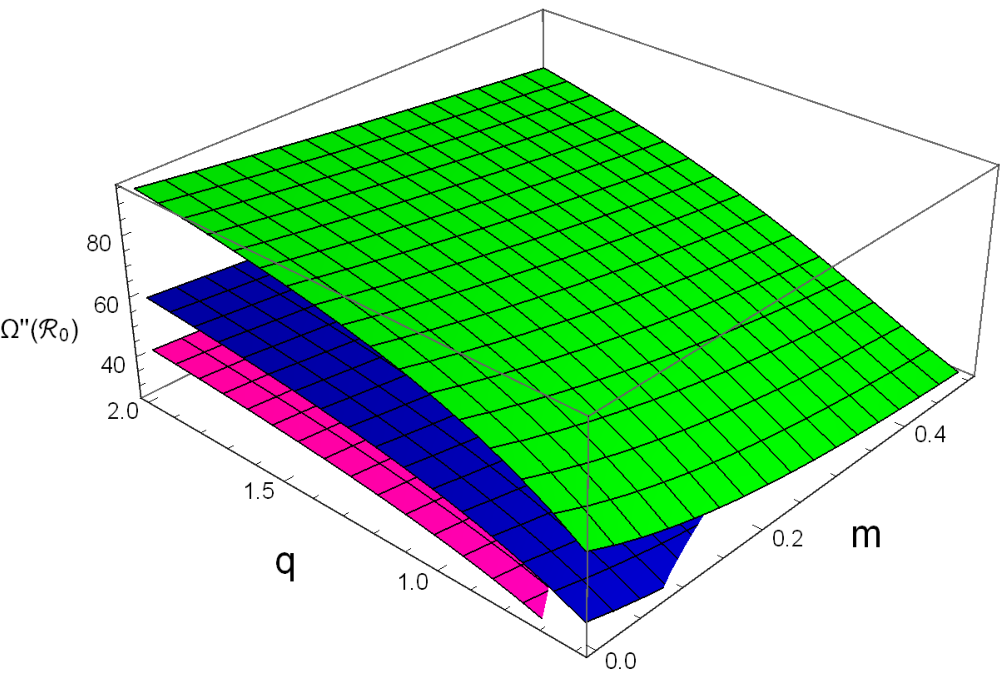,width=.5\linewidth}
\caption{Stability of thin-shell WH with phantomlike EoS for
different values of $n$ as $n=-1$(magenta) $n=-2$(blue)
$n=-5$(green) with $m=0.1$ (left plot) and $\alpha=-2$ (right
plot).}
\end{figure}
\begin{figure}\centering
\epsfig{file=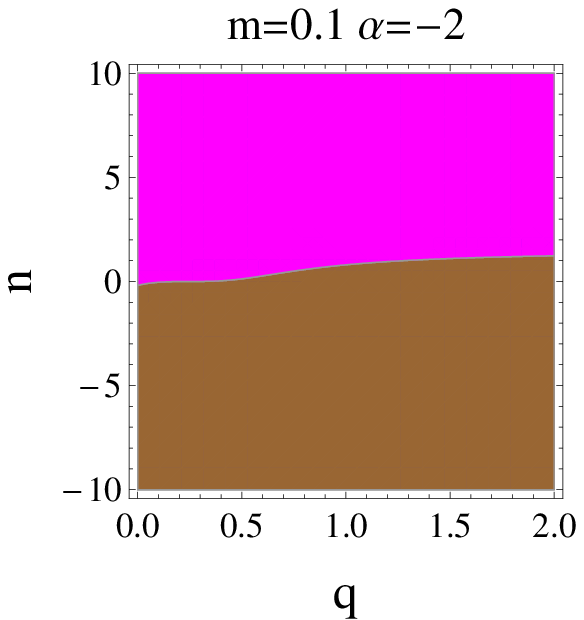,width=.325\linewidth}\epsfig{file=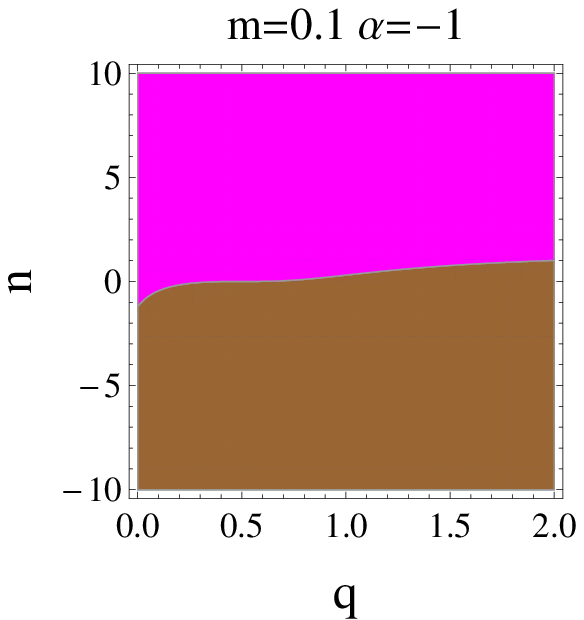,width=.325\linewidth}\epsfig{file=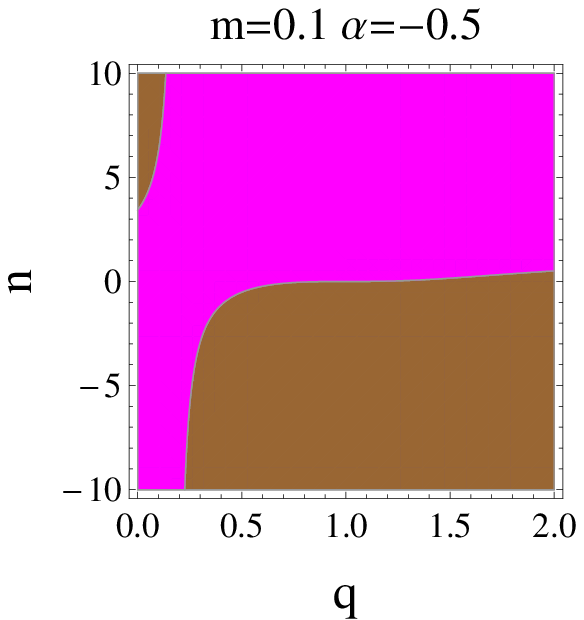,width=.325\linewidth}
\caption{Region plots of $\Omega''(\Re_{0})$ verses $q$ and $n$ for
Chaplygin variable EoS with different values of $\alpha$.}
\end{figure}

\subsection{Chaplygin Variable EoS}

In the last case, we consider Chaplygin variable EoS written as
\cite{56}
\begin{equation}\label{40}
\mathcal{P}=\frac{1}{\Re^n}\frac{C}{\rho},
\end{equation}
where $C$ is the EoS parameter. This reduces to Chaplygin EoS for
$n\rightarrow0$. The solution of conservation equation in terms of
surface energy density is
\begin{figure}\centering
\epsfig{file=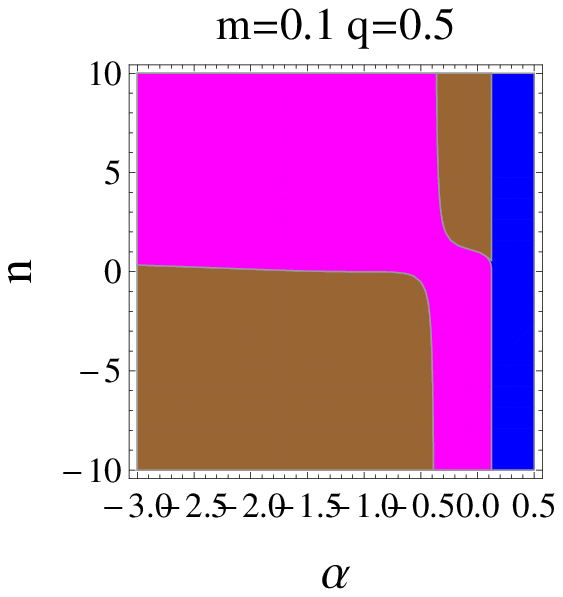,width=.325\linewidth}\epsfig{file=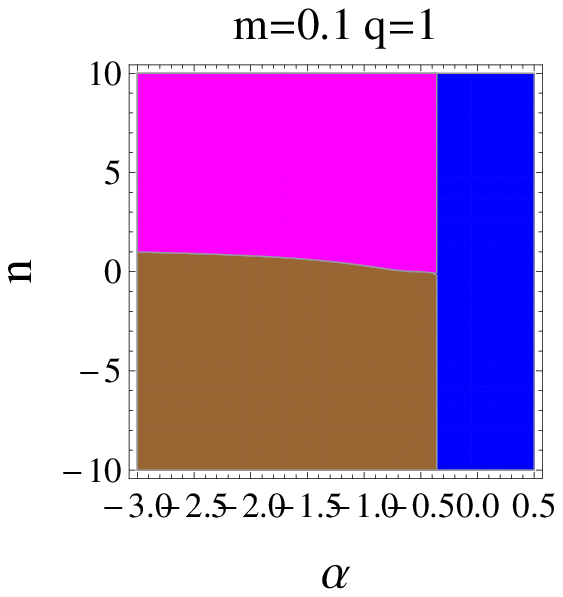,width=.325\linewidth}\epsfig{file=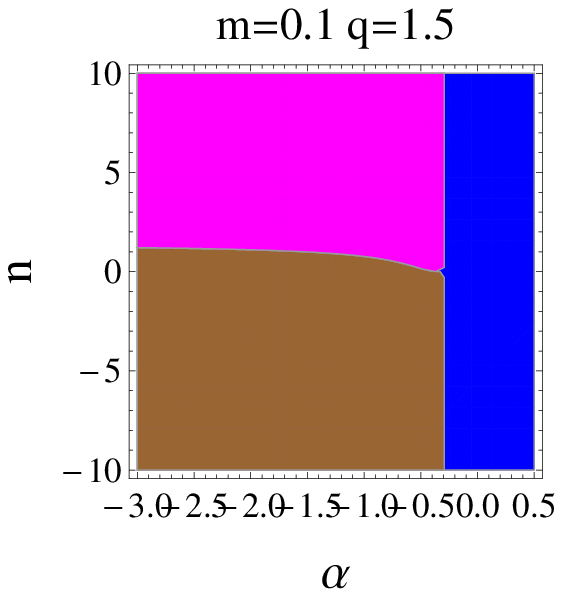,width=.325\linewidth}
\caption{Region plots of $\Omega''(\Re_{0})$ verses $\alpha$ and $n$
for Chaplygin variable EoS with different values of $q$.}
\epsfig{file=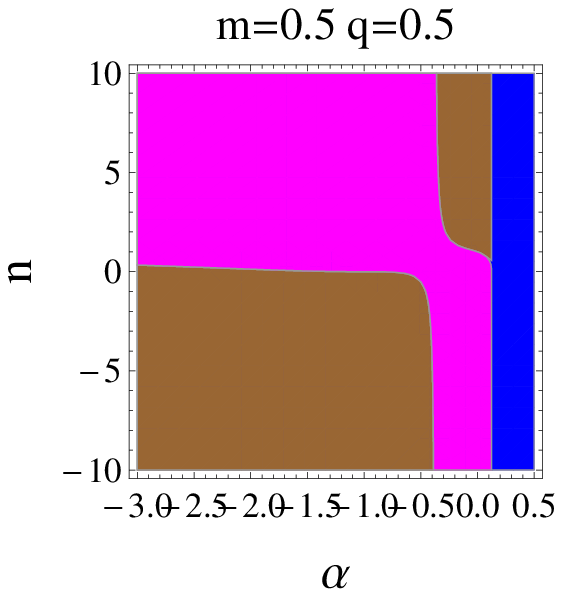,width=.325\linewidth}\epsfig{file=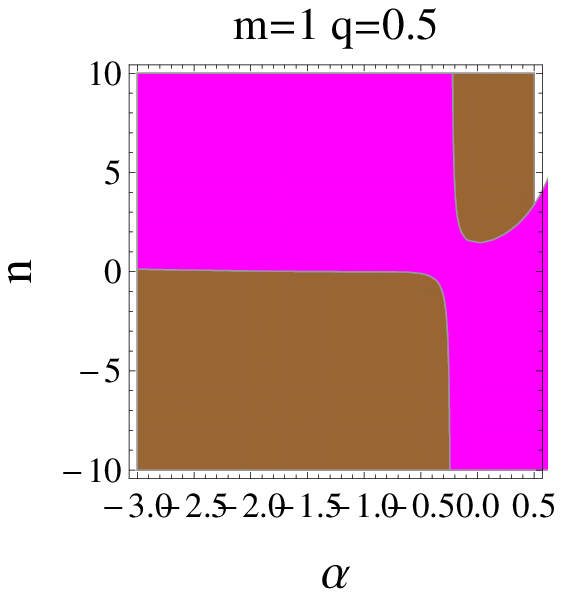,width=.325\linewidth}\epsfig{file=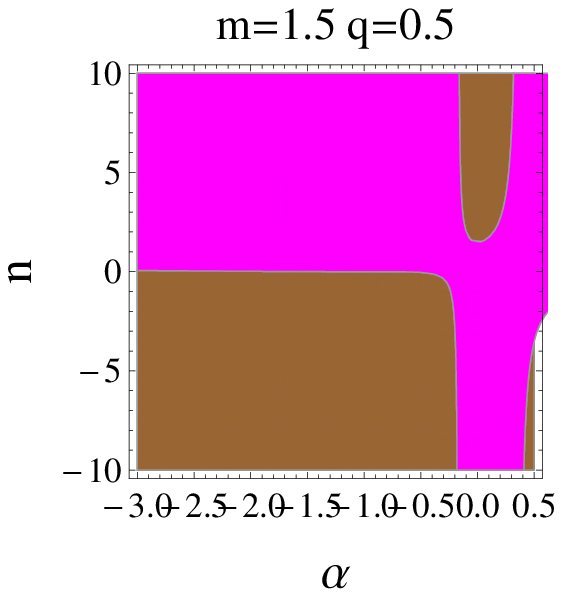,width=.325\linewidth}
\caption{Region plots of $\Omega''(\Re_{0})$ verses $\alpha$ and $n$
for Chaplygin variable EoS with different values of $m$.}
\epsfig{file=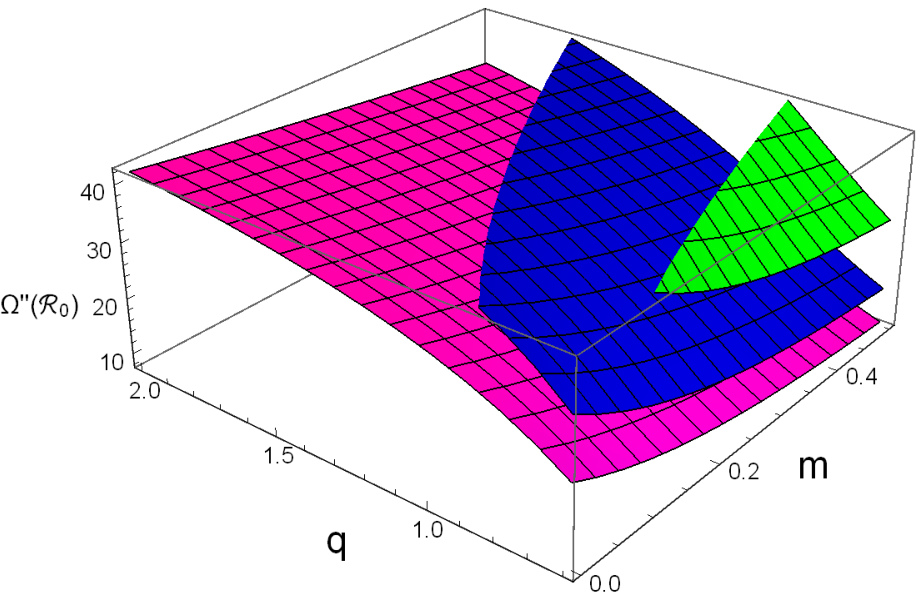,width=.5\linewidth}\epsfig{file=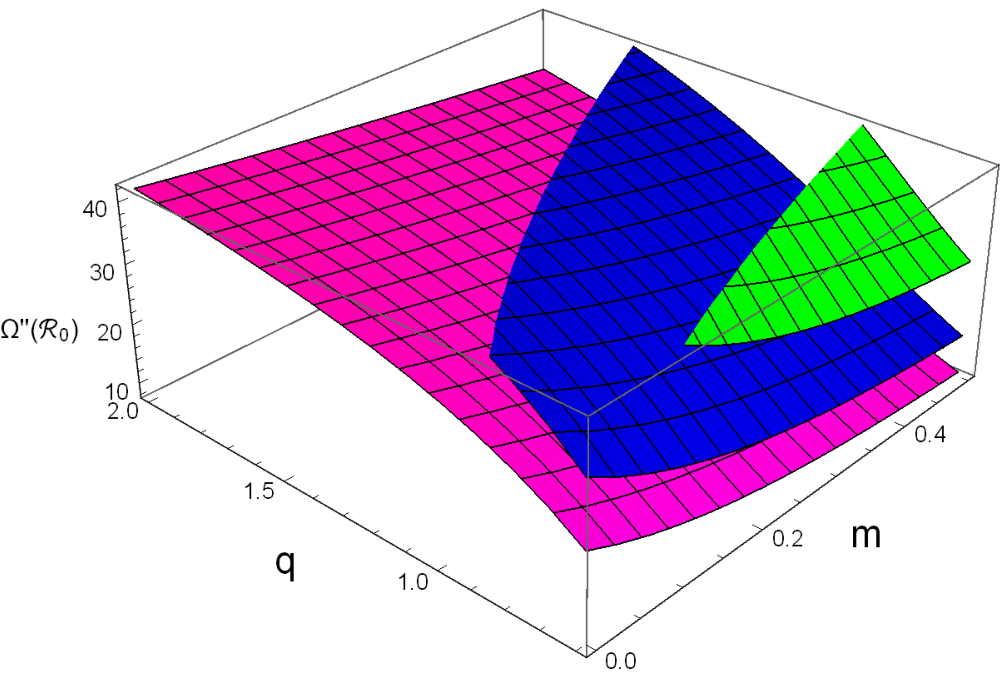,width=.5\linewidth}
\caption{Stability of thin-shell WH with Chaplygin gas for different
values of $n$ as $n=-1$(magenta) $n=-2$(blue) $n=-5$(green) with
$m=0.1$ (left plot) and $\alpha=-2$ (right plot).}
\end{figure}
\begin{equation}\label{41}
\rho^2(\Re)=\frac{4C\left(\Re^4\Re_0^n-\Re^n\Re_0^4\right)+
\rho_{0}^2\Re_0^{n+4}\Re^n(n-4)}{(n-4)\Re^{n+4}\Re_0^{n}}.
\end{equation}
We obtain the respective potential function for such type of matter
distribution and noted that $\Omega(\Re_0)=0$. Also, we calculate
$\Omega'(\Re)$ and taking $\Omega'(\Re_0)=0$ so that
\begin{equation}\label{42}
C=-\frac{m \Re_0^{n-3}}{8 \pi ^2}+\frac{q^2 \Re_0^{n-4}}{8 \pi
^2}+\frac{\alpha ^2 \Re_0^n}{24 \pi ^2}-\frac{1}{2} \rho _0^2
\Re_0^n.
\end{equation}
Second derivative of potential function with respect to shell radius
at $\Re=\Re_0$ written as
\begin{eqnarray}\label{43}
\Omega''(\Re_0)&=&-\frac{2 m n}{\Re_0^3}+\frac{6
m}{\Re_0^3}-\frac{1}{3} 4 \alpha ^2 n+\frac{4 \alpha  n
q}{\Re_0^2}+\frac{2 n}{\Re_0^2}-\frac{8 \alpha
q}{\Re_0^2}-\frac{4}{\Re_0^2}.
\end{eqnarray}
For Chaplygin variable EoS, the maximum stable regions are obtained
only if both $n$ and $\alpha$ are negative as shown in Figs. 11-13.
It is found that stable regions increase for negative values of
$\alpha$ and decreases as $\alpha$ approaches positive values.
Higher values of charge also enhance the stable regions. Similarly,
we have obtained that the stability of the developed structure is
maximum as $n$ as well as $\alpha$ have maximum negative values and
stability decreases as $\alpha$ or $n$ approaches zero or positive
values (Fig. 14).

\section{Concluding Remarks}

This paper has explored the effects of nonlinear electrodynamics on
the stable configuration of thin-shell WHs. For this purpose, we
have constructed WH geometry through the matching of two equivalent
copies of RN BH with nonlinear electrodynamics. The components of
matter filled at thin-shell are found by using a reduced form of
Einstein field equations known as Lanczos equations. The developed
structure is physically acceptable for the WH geometry as null and
weak energy conditions are violated (Fig. 1). It is noted that the
attractive and repulsive nature of the WH throat is greatly affected
by the coupling parameter $\alpha$ (Fig. 2). We have analyzed the
stability of the developed structure by using radial perturbation
with three different types of matter distributions, i.e.,
barotropic, phantomlike variable, and Chaplygin variable EoS.

Firstly, we have considered barotropic type fluid distribution and
analyzed the stable configuration graphically (Figs. 3-6). It is
worthwhile to mention that the developed structure shows stable
regions for barotropic EoS if $\alpha<0$ (Figs. 3-5). It is noted
that WH geometry developed from Schwarzschild and RN BHs express
unstable behavior for every choice of physical parameter for
barotropic EoS \cite{56,57}. Hence, the presence of nonlinear
electrodynamics provides stability of WH geometry for barotropic
EoS. It is found that stable regions increase by an increasing
charge of the geometry with negative values of coupling parameters.
These results also supported the final results of published articles
\cite{56}-\cite{59}.

Secondly, we have interested to observe the stability of thin-shell
filled with phantomlike variable EoS (Figs. 7-10). It is found that
stable regions must exist for every choice of the physical parameter
if $n<0$ and $\alpha<0$ (Figs. 7-9). There is also a possibility for
a stable structure if both $\alpha$ and $n$ are positive. It is also
noted that the developed structure is more stable in the presence of
nonlinear electrodynamics. As thin-shell, WHs developed from
Schwarzschild, RN, Bardeen, and Bardeen-de Sitter BHs have less
stable configurations for some specific values of the physical
parameter while RN BH with nonlinear electrodynamics is more
suitable for the construction of stable thin-shell WH
\cite{56}-\cite{59}.

Finally, we have analyzed the effects of Chaplygin variable EoS on
the stability of the developed structure (Figs. 11-14). It is found
that there exist more stable regions if both $\alpha$ and $n$ have
the same sign (Figs. 11-13). It is also noted that the stability of
the geometry increases by an increasing charge of the geometry. The
stable configuration decreases as $\alpha$ and $n$ lead to positive
values and increases as $\alpha$ and $n$ approach large negative
values (Fig. 14).

It is concluded that thin-shell WH with nonlinear electrodynamics
expresses more stable regions as compared to WH geometry developed
from Schwarzschild, RN, Bardeen and Bardeen-de Sitter BHs
\cite{56}-\cite{59}.

\end{document}